# ONTOLOGY FOR CELLULAR COMMUNICATION

HASNI, Neji[1]

[1]6'Tel Research Unit, Higher School of Communications of Tunis, Sup'Com, Tunisia

hasni.neji63@laposte.net

**ABSTRACT**

The lack of interoperability between mobile cellular access networks has long been a challenging obstacle, which telecommunication engineering is trying to overcome. In second generation networks for example, this problem lies in the fact that there are multiple standards. Each of these standards can operate in the same frequency range. However, each utilizes a different Radio Technology and Modulation Scheme, which are characteristics of the standard. Therefore, the lack of interoperability in 2G occurs because of the lack of standardization.
Interoperability within 3G networks is limited to a few operating modes using different Radio Transmission Technologies that are not inter-operable. Thus, interoperability remains an issue for 3G. 4G technology even being successful in its various trials cannot guarantee the interoperability. This is within each network generation; meanwhile between heterogeneous network generations the situation seems to be worst. The undertaken approach to overcome the interoperability issues between heterogeneous network generations begins by establishing a holistic understanding of cellular communication systems (a starting point for interoperability improvement), and proposing an ontological approach that expresses the cellular communication systems' concepts, classes, and properties in a formal and unambiguous way.
This approach is first to analyze the structure, inputs, and outputs of three different cellular technologies, performing a domain analysis (of this subset of technologies) and producing a feature model of the domain. Finally, we sought to build an ontology capable of providing a common view of the domain, providing an effective representation of relations between representations of corresponding concepts in different cellular technologies.

## 1. Introduction

Developing an environment that maximizes interoperability, communication and efficiency tailored for particular domains is a common objective for researchers; who seek to improve the outputs, by automating engineering practice around a specific domain.
The advantage of developing specific ontologies tailored to the telecommunication's domain provides benefits stemming from representational efficiency. However, there has not been a lot of work in developing ontologies tailored to the domain of wireless cellular communication itself. One reason for this is the amount of effort required to produce such ontology is substantial. Specific ontologies such as this ongoing one are, in fact, not easily buildable, which will require from us to undertake seemingly heavy processes to identify existing features in three different wireless cellular networks technologies to satisfy the representational needs. An ideal solution will be offered by the construction of a general ontology for common features management, which might allow for resource sharing over and across multiple technologies in the telecommunication domain, possibly with an easy and fast process of customization without having to develop new systems from scratch.

## 2. Application of Ontologies for Interoperability

### 2.1 Ontology overview

The term "Ontology" is borrowed from philosophy where it is defined as a systematic investigation of "Existence". The term is now widely used in Artificial Intelligence and Knowledge Engineering where what "exists" are those entities which can be "represented." Ontology is the term used to refer to the shared understanding of some domain of interest that may be used as a unifying framework to solve problems in that domain [USCH96].

Ontology necessarily entails or embodies a world view with respect to a given domain. This world view is often conceived as a set of concepts (e.g. entities, attributes, processes) along with their definitions and their inter-relationships.

Because people, organizations, and cellular technologies must communicate between and among themselves, there are often difficulties/inaccuracies in communications because of differing contexts, understandings, viewpoints and assumptions. Therefore, ontologies help to accomplish the following:
• Improve poor communication,
• Establish a unifying framework for conceptual models and ideas,
• Establish the basis for interoperability, and
• Prevent redundant work and cross purposes.

### 2.2 A proposed development strategy

Our strategy for developing the ontology will be based on both a top-down and bottom-up approach. In order to be effective, we sought to make the top-down approach tackle the core problem of the interoperability between the cellular network's technologies. The bottom-up approach will focus on developing specific cellular network's technologies ontologies that accurately described the artifacts produced by the cellular network's technologies, so that their data processes could be actually made to interoperate.

A cellular network's technologies ontology is a system of features, selected because of their usefulness to capture interesting commonalities and similarities between technologies. The choice of a proper ontology for the cellular network's technologies is a very important factor in accomplishing the task of interoperability building and structuring, far beyond the issue of the representation of the inventory of the cellular networks' features.

### 2.3 Methodology

Because there is currently no ontology for the domain of cellular network's technologies, we are unable to rely on previous work and instead have to develop our own ontology. We are, however, able to leverage an existing methodology for establishing our ontology and tailor that methodology to our purpose.

The ontology development process starts by identifying the purpose and scope of the ontology (step 1).

The second step (step 2) is the development of feature analysis for the selected domain (in this case, the domain of cellular network's technologies).

This is followed by (step 3) reasoning and brainstorming about observations and information generated by the feature models to select the commonalities between the three cellular network's technologies and build a high level ontology representing these commonalities.

The next step (step 4) is to build more detailed ontologies for each technology. These ontologies include more essential characteristics at a finer level of granularity.

Next (step 5), we used UML [CRAN99] to represent the relationships between the three ontologies.

Finally, we documented the ontology (step 6).

### 2. 4 Feature Modeling [CZAR00]

To perform a domain analysis of the subset of cellular technologies, we will proceed by producing a feature model for each technology of the domain of interest.

Features are originally used to define software product lines and system families, to identify and manage commonalities and variabilities between products and systems. Attempting to define a feature model for existing cellular network's technologies allows us to explore, identify, and define the key aspects of existing technologies so that these aspects can be described in ontology. It is this ontology that then allows us to improve interoperability between existing cellular network's technologies.

Our approach for the analysis and the investigation of the structure of inputs, outputs, and relationships of a collection of individual cellular technologies can be characterized as a domain analysis (of this subset of technologies) and the production of feature model of that domain. This technique is well suited for the cellular technologies' features as well as the identification of their essential characteristics. Use of these characteristics in further steps of the research allows them to interoperate.

The feature model is an abstract representation of functionality found in the domain. It is used during domain engineering in order to obtain an abstract view on this functionality, which can be verified against the needs raised by the domain. Therefore, each feature is a relevant characteristic of the domain.

The description of feature models was tied to the introduction of the Feature-Oriented Domain Analysis (FODA[1]) approach in the late eighties. A feature model represents an explicit model of a device or system by summarizing the features and the variation points of the device/system. Feature models include the rationale (a feature should have a note explaining why the feature is included in the model) and the stakeholders for each of feature. A feature model for software system captures the reusability and configurability aspects of reusable software. They feature model also provides a road map to variability in other models (e.g. object models, use case models, interaction and state transition diagram). Griss et al. describes the important relationship between use case models and feature models as follows [CZAR00]: a use case model captures the system requirements from the user perspective (operational requirements), whereas the feature model organizes requirements from the user perspective based on commonality and availability analysis.

## 3. Contributions

The telecommunication contributions that will be represented in this field are:
- An initial investigation and analysis of the structure, inputs, and outputs of three cellular network technologies, and the identification of essential characteristics of these technologies.
- The completion of a domain analysis (of this subset of technologies) and production of a feature model for each technology's characteristics.
- An identification of the commonalities between the three cellular network's technologies characteristics that must be accounted for in building a high level ontology for the domain.
- The construction of an initial high-level ontology using a knowledge based design and knowledge system developed at Stanford University: "Protégé 2000" [PROT20] that can be used in any field where the concepts can be modeled as a class hierarchy.
- Because Protégé-2000 is designed to support *iterative development*, where there are cycles of revision to the ontologies and other components of the knowledge-based system. We propose an establishment of a methodology around which other cellular network technologies could be analyzed and added to this initial cellular network's technologies ontology.

## 4. Conclusion

While there has been plenty of research into the development of single aspects in the mobile cellular communication field, and some specific frameworks developments, there has been little research into holistic models to define how these various threads and processes could (and should) most efficiently and effectively interact smoothly. The development of a holistic framework potentially provides seamless interoperability between cellular network's technologies allowing them to communicate more efficiently and reliably with high quality. Additionally, the existence of such interoperability enhances the discovery of dependencies among different aspects of the cellular network's technologies. The hope is that it will enable telecommunication researchers to discover various improvements. The long-term goal of this research is to support all aspects of cellular technologies; however, the immediate goal is to demonstrate the theoretical feasibility of integrating a selected subset of cellular network's technologies using ontologies

---

[1] Feature-oriented domain analysis (FODA) is a domain analysis method developed at the Software Engineering Institute (SEI). The method is known for the introduction of feature models and feature modeling.

## 5. Acknowledgements

It would have been impossible for me to undertake an endeavor of this size and scope without the assistance of my theses supervisor, professor BOUALLEGUE, Ridha. It is with sincerest thanks that I would like to acknowledge his help, contribution, and guidance t*o* my education. His tireless efforts to steer me in the right directions, clarify my understanding of complex material, and nudged me when I needed nudging.

## 6. References

[CRAN99]  Cranefield, S. and Purvis, M., "UML as an Ontology Modelling Language", *Proceedings of the IJCAI'99 Workshop on Intelligent Information Integration*, Sweden, 1999.

[CZAR00]  Czarnecki, K. and Eisenecker, U., *Generative Programming Methods, Tools, and Applications*, Addison-Wesley, 2000.

[PROT20]  Protégé, [http://protege.stanford.edu], 20 October 2002.

[USCH96]  Uschold, M. and Gruninger, M., "Ontologies: Principles, Methods and Applications," *Knowledge Engineering Review*, Vol. 11, No. 2, June 1996.